\documentclass[twocolumn, twocolappendix]{aastex63} 

\usepackage{epsfig,natbib}
\usepackage{graphicx}
\usepackage{amsmath}
\usepackage{booktabs}
\usepackage{xspace}
\usepackage{hyperref}
\usepackage[T1]{fontenc}
\usepackage{lipsum}
\usepackage{bm}
\usepackage{xcolor}
\newcommand{\mearth}{$M_\earth$~}
\newcommand{\mearthe}{$M_\earth$}

\newcommand{\msini}{\ensuremath{M \sin i}}

\newcommand{\fe}{[Fe/H]\xspace}

\newcommand{\ms}{m s$^{-1}$\xspace}

\newcommand{\Msun}{\ensuremath{M_{\odot}}\xspace}
 
\newcommand{\K}{\ensuremath{L_{\odot}}\xspace}

\shortauthors{Fulton et al.}
\shorttitle{California Legacy Survey II: Giant Planet Occurrence}
\begin{document}

\title{California Legacy Survey II. Occurrence of Giant Planets Beyond the Ice line}

\author[0000-0003-3504-5316]{Benjamin J.\ Fulton}
\affiliation{Cahill Center for Astronomy $\&$ Astrophysics, California Institute of Technology, Pasadena, CA 91125, USA}
\affiliation{IPAC-NASA Exoplanet Science Institute, Pasadena, CA 91125, USA}

\author[0000-0001-8391-5182]{Lee J.\ Rosenthal}
\affiliation{Cahill Center for Astronomy $\&$ Astrophysics, California Institute of Technology, Pasadena, CA 91125, USA}

\author[0000-0001-8058-7443]{Lea A.\ Hirsch}
\affiliation{Kavli Institute for Particle Astrophysics and Cosmology, Stanford University, Stanford, CA 94305, USA}

\author[0000-0002-0531-1073]{Howard  Isaacson}
\affiliation{Department of Astronomy, University of California Berkeley, Berkeley, CA 94720, USA}
\affiliation{Centre for Astrophysics, University of Southern Queensland, Toowoomba, QLD, Australia}

\author[0000-0001-8638-0320]{Andrew W.\ Howard}
\affiliation{Cahill Center for Astronomy $\&$ Astrophysics, California Institute of Technology, Pasadena, CA 91125, USA}

\author[0000-0001-9408-8848]{Cayla M.\ Dedrick}
\affiliation{Cahill Center for Astronomy \& Astrophysics, California Institute of Technology, Pasadena, CA 91125, USA}
\affiliation{Department of Astronomy \& Astrophysics, The Pennsylvania State University, 525 Davey Lab, University Park, PA 16802, USA}

\author[0000-0001-7730-0202]{Ilya A.\ Sherstyuk}
\affiliation{Cahill Center for Astronomy $\&$ Astrophysics, California Institute of Technology, Pasadena, CA 91125, USA}

\author[0000-0002-3199-2888]{Sarah C.\ Blunt}
\affiliation{Cahill Center for Astronomy $\&$ Astrophysics, California Institute of Technology, Pasadena, CA 91125, USA}
\affiliation{NSF Graduate Research Fellow} 

\author[0000-0003-0967-2893]{Erik A.\ Petigura}
\affiliation{Department of Physics $\&$ Astronomy, University of California Los Angeles, Los Angeles, CA 90095, USA}

\author[0000-0002-5375-4725]{Heather A.\ Knutson}
\affiliation{Division of Geological and Planetary Sciences, California Institute of Technology, Pasadena, CA 91125, USA}


\author[0000-0003-0012-9093]{Aida Behmard}
\affiliation{Division of Geological and Planetary Sciences, California Institute of Technology, Pasadena, CA 91125, USA}
\affiliation{NSF Graduate Research Fellow} 

\author[0000-0003-1125-2564]{Ashley Chontos}
\affiliation{Institute for Astronomy, University of Hawai$'$i, Honolulu, HI 96822, USA}
\affiliation{NSF Graduate Research Fellow} 

\author[0000-0003-0800-0593]{Justin R.\ Crepp}
\affiliation{Department of Physics, University of Notre Dame, Notre Dame, IN, 46556, USA}

\author[0000-0002-1835-1891]{Ian J.\ M.\ Crossfield}
\affiliation{Department of Physics and Astronomy, University of Kansas, Lawrence, KS, USA}

\author[0000-0002-4297-5506]{Paul A.\ Dalba}
\affiliation{Department of Earth and Planetary Sciences, University of California, Riverside, CA 92521, USA}
\affiliation{NSF Astronomy $\&$ Astrophysics Postdoctoral Fellow}

\author[0000-0003-2221-0861]{Debra A.\ Fischer}
\affiliation{Department of Astronomy, Yale University, New Haven, CT 06511, USA}

\author[0000-0003-4155-8513]{Gregory W.\ Henry} 
\affiliation{Center of Excellence in Information Systems, Tennessee State University, Nashville, TN 37209 USA}

\author[0000-0002-7084-0529]{Stephen R.\ Kane}
\affiliation{Department of Earth and Planetary Sciences, University of California, Riverside, CA 92521, USA}

\author[0000-0002-6115-4359]{Molly Kosiarek}
\affiliation{Department of Astronomy and Astrophysics, University of California, Santa Cruz, CA 95064, USA}
\affiliation{NSF Graduate Research Fellow}

\author[0000-0002-2909-0113]{Geoffrey W. Marcy}
\affiliation{Department of Astronomy, University of California Berkeley, Berkeley, CA 94720, USA}

\author[0000-0003-3856-3143]{Ryan A.\ Rubenzahl}
\affiliation{Cahill Center for Astronomy $\&$ Astrophysics, California Institute of Technology, Pasadena, CA 91125, USA}
\affiliation{NSF Graduate Research Fellow} 

\author[0000-0002-3725-3058]{Lauren M.\ Weiss}
\affiliation{Institute for Astronomy, University of Hawai$'$i, Honolulu, HI 96822, USA}

\author[0000-0001-6160-5888]{Jason T.\ Wright}
\affiliation{Department of Astronomy and Astrophysics, The Pennsylvania State University, University Park, PA 16802, USA}
\affiliation{Center for Exoplanets and Habitable Worlds, The Pennsylvania State University, University Park, PA 16802, USA}
\affiliation{Penn State Extraterrestrial Intelligence Center, The Pennsylvania State University, University Park, PA 16802, USA}

\begin{abstract}

We used high-precision radial velocity measurements of FGKM stars to determine the occurrence of giant planets as a function of orbital separation spanning 0.03--30 au. Giant planets are more prevalent at orbital distances of 1--10 au compared to orbits interior or exterior of this range.  The increase in planet occurrence at $\sim$1 au by a factor of $\sim$4 is highly statistically significant.  A fall-off in giant planet occurrence at larger orbital distances is favored over models with flat or increasing occurrence.  We measure $14.1^{+2.0}_{-1.8}$ giant planets per 100 stars with semi-major axes of 2--8 au and $8.9^{+3.0}_{-2.4}$ giant planets per 100 stars in the range 8--32 au, a decrease in giant planet occurrence with increasing orbital separation that is significant at the $\sim$2$\sigma$ level. We find that the occurrence rate of sub-Jovian planets (0.1--1 Jupiter masses) is also enhanced for 1--10 au orbits. This suggests that lower mass planets may share the formation or migration mechanisms that drive the increased prevalence near the water-ice line for their Jovian counterparts. Our measurements of cold gas giant occurrence are consistent with the latest results from direct imaging surveys and gravitational lensing surveys despite different stellar samples. We corroborate previous findings that giant planet occurrence increases with stellar mass and metallicity.
\end{abstract}


\section{Introduction}
\label{sec:intro}

Expanding and characterizing the population of known exoplanets with measured masses and orbital periods is crucial to painting a more complete picture of planet formation and evolution. A census of diverse exoplanets sheds light on worlds radically different from Earth and can provide insight into how these planets---and those orbiting the Sun---formed. Ground-based radial velocity (RV) surveys measure the Doppler shifts of stellar spectra to discover exoplanets and characterize their orbits and masses.  These surveys have provided landmark discoveries that shaped our understanding of the formation and architectures of other worlds \citep[e.g.,][]{Mayor95,Marcy02,Tamuz08}. 

Doppler planet searches take time to accumulate the time series measurements that trace out planetary orbits.  The Keck Planet Survey \citep{Cumming08} used eight years of RVs from Keck-HIRES \citep{Vogt94} to make the first broad measurement of giant planet occurrence ($\msini \geq 0.1  M_J$). This survey discovered an increase in the abundance of giant planets for orbits near the water-ice line and found that about 10\% of Sun-like stars have giant planets with a semi-major axes of $<$3 au.  The survey only reported planet detections for orbital periods shorter than 2000 days, the observational baseline of the survey.  Extrapolating based on the detection of partial orbits, \cite{Cumming08} estimated that $\sim$20\% of such stars have a giant planet orbiting within 20 au.

Other teams of astronomers have surveyed the Northern and Southern skies in parallel with the Keck search.  \cite{Mayor11} used 8 years of precise HARPS RVs supplemented by additional RVs from CORALIE to measure occurrence patterns in the population of giant planets that are similar to those described above.  They found that the planet mass function is ``bottom heavy''. That is, low-mass planets (0.3--30 \mearth) are significantly more common than giant planets, a finding consistent with measurements from Keck Observatory by \cite{Howard10_Science}. Since then, the HARPS team has continued to discover increasingly longer-period and lower-mass planets \citep{Udry17, Rickman19}.  Two other `legacy' planet searches have contributed significantly to our knowledge of giant planets.  \cite{Wittenmyer20} used data from a subset of the stars surveyed by the 18-year Anglo-Australian Planet Search, which has also uncovered a number of cold giant planets \citep{Wittenmyer17, Kane19}, to measure a significant increase in giant planet occurrence at $\sim$1 au and a constant occurrence for orbits in the range $\sim$1--6 au. Similarly, the McDonald Observatory planet search has been operating for more than 20 years using the 2.7-m Harlan J. Smith Telescope, and has contributed valuable discoveries of long-period giant planets \citep[e.g.,][]{Robertson2012, Endl2016, Blunt19}.

We are now in the fourth decade of Doppler planet searches. As we begin to discover planets with orbital periods comparable to Saturn's, we can answer questions that require a rigorous accounting of giant planets spanning a large range of orbital distances. What is the mass versus semi-major axis distribution of planets out to 10 au? How abundant are cold gas giants beyond the water-ice line, and what can this abundance tell us about planet formation across protoplanetary disks? 

The California Legacy Survey (CLS, Rosenthal et al. 2021) is uniquely suited for this work. As an unbiased radial velocity survey of 719 stars over three decades, the CLS is an excellent sample for a variety of occurrence measurements, particularly for cold gas giants. In this paper, we explore giant planet occurrence as a function of orbital separation. In Section 2, we review the star and planet catalog of the California Legacy Survey. In Section 3, we describe our methods for computing planet occurrence. Section 4 describes the patterns of planet occurrence that we observe in the sample. In Section 5, we discuss our findings and their context. We summarize our work in Section 6.

\section{Survey Review}
\label{sec:survey}

The California Legacy Survey is a Doppler search for planets orbiting a well-defined sample of nearby FGKM stars conducted by the California Planet Search team \citep[CPS;][]{Howard10}.  Paper I in this series (Rosenthal et al. 2021) describes the CLS in detail, including the stellar sample, the search methodology, and the resulting planet sample upon which this paper and forthcoming works in the CLS paper series build.  The CLS stellar sample was selected specifically to make the measurements reported here---planet occurrence measurements, especially of giant planets with orbits out to 10 au and beyond---and it approximates a random sample of nearby stars. In particular, stars were selected for CLS observations independent of whether planets were known to orbit them.  Stars were also selected independent of their metallicity or other factors that might make them more or less likely to harbor planets.

CLS builds on Doppler measurements from the Keck Planet Search \citep{Cumming08}, a touchstone Doppler survey of 585 stars observed with HIRES at the W.\,M.\ Keck Observatory during 1996--2004.  We continued to observe those stars and an additional 134 stars at Keck Observatory through 2020.  CLS also includes observations of a subset of these stars made with the Hamilton spectrometer at Lick Observatory during 1988--2011, high-cadence Keck-HIRES observations of 235 magnetically inactive stars as part of the Eta-Earth Survey \citep{Howard10_Science}, and high-cadence Lick-APF observations of 135 of those stars \citep{Fulton16, Hirsch21}.  The average star has been observed for 22 years and has 71 RVs with a precision of $\sim$2\,\ms. While these stars do not have homogeneous observing histories, our search methodology accounts for this by incorporating the search completeness of each star's individual dataset.  (A Doppler survey that is completely homogeneous in the number, precision, and temporal spacing of measurements is infeasible given the three decade history of this planet search---indeed, this survey spans an era longer than the time during which extrasolar planets orbiting Sun-like stars have been known!)  By the metric of ``Doppler survey \'etendue'' (number of stars surveyed $\times$ typical time series duration), CLS is the largest planet search to date at the $\sim$\ms\ level.

Our search methodology (described in Rosenthal et al. 2021) involves an automated, iterative periodogram-based search for Keplerian signals with uniform vetting to identify false positives. This methodology detected 177 planets orbiting the 719 stars in the CLS stellar sample. The algorithm is sensitive to orbital periods much longer than the baseline of our dataset, with the longest period signals detected as partial orbits. 

The search was also sensitive to orbital segments only seen as linear and parabolic trends in an RV time series.  There were only six such detections in our sample of trends that are not associated with known stellar binaries and are potentially consistent with planetary mass companions.  Thus, nearly all orbital signals were resolved or partially resolved as Keplerian signals.

\begin{figure*}[ht!]
\begin{center}
\includegraphics[width = \textwidth]{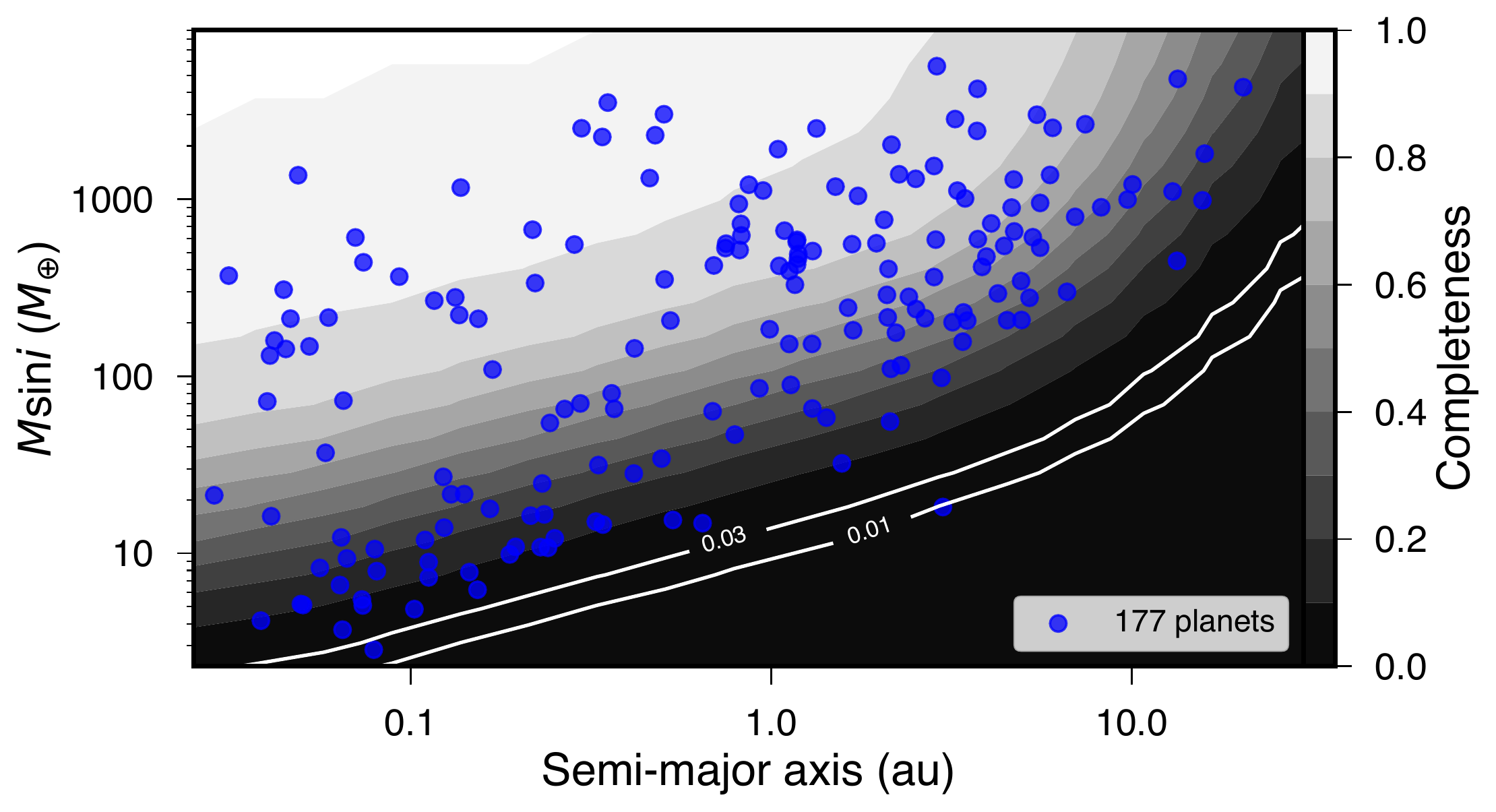}
\caption{California Legacy Survey planet catalog and survey-averaged search completeness contours in semi-major axis and \msini . 3$\%$ and 1$\%$ search completeness contours are highlighted in white.}
\label{fig:sample}
\end{center}
\end{figure*}

To characterize survey completeness for each star in the survey, we conducted injection-recovery tests of synthetic Doppler planet signals over a range of injected masses, orbital periods, and orbital geometries.  Detected planets and CLS survey completeness are shown in Figure \ref{fig:sample}. We refer the reader to Rosenthal et al. (2021) for the full stellar sample and planet catalog.

The CLS stellar sample has a median metallicity of \fe = 0.0\,dex, a median stellar mass of 1.0\,\Msun, and a small number of evolved stars (subgiants). These are good heuristics for verifying that we successfully constructed a blind occurrence survey, since a bias toward known giant planet hosts could manifest as a metal-rich sample \citep{Fischer05, Santos04a}, a particularly massive sample, or an excess of evolved stars \citep{Johnson11}.

\section{Methods}
\label{sec:occurrence}

The primary goal of this work is to measure planet occurrence. Many studies of RV or transit surveys use the intuitive occurrence measurement method known as ``inverse detection efficiency'' \citep{Howard12, Petigura13b}. According to this procedure, one estimates occurrence in a region of parameter space by counting the planets found in that region, with each planet weighted by the local search completeness. One can measure the search completeness map of a survey by injecting many synthetic signals into each dataset and computing the fraction of signals in a given region that are recovered by the search algorithm in use. Inverse detection efficiency is actually a specific case of a Poisson likelihood method, in which one models an observed planet catalog as the product of an underlying Poisson process and empirical completeness map \citep{Foreman-Mackey14}. This can be done with a parametric occurrence rate density model, like a broken power law, or a non-parametric density model, with a piecewise-constant step function. In this paper, we used the Poisson likelihood method to model the occurrence of giant planets, taking measurement uncertainty into account.

We used the hierarchical Bayesian methodology outlined in \cite{Hogg10} and \cite{Foreman-Mackey14} to evaluate our occurrence likelihood. Given an observed population of planets with orbital and \msini\ posteriors $\{\bm{\omega}\}$ and associated survey completeness map $Q(\bm{\omega})$, and assuming that our observed planet catalog is generated by a set of independent Poisson process draws, we evaluated a Poisson likelihood for a given occurrence model $\Gamma(\bm{\omega} | \bm{\theta})$, where $\Gamma$ is an occurrence density $\frac{\mathrm{d}^2N}{\mathrm{dln}(a)\mathrm{dln}(\msini)}$ and $\bm{{\theta}}$ is a vector of model parameters. The observed occurrence $\hat{\Gamma}(\bm{\omega} | \bm{\theta})$ of planets in our survey can be modeled as the product of the measured survey completeness and an underlying occurrence model,

\begin{equation}
\hat{\Gamma}(\bm{\omega} | \bm{\theta}) = Q(\bm{\omega})\Gamma(\bm{\omega} | \bm{\theta}). \\
\end{equation}

The Poisson likelihood for an observed population of objects is

\begin{equation}
\mathcal{L} = e^{-\int \hat{\Gamma}(\bm{\omega} | \bm{\theta}) \,d\bm{\omega}} \prod_{k=1}^{K} \hat{\Gamma}(\bm{\omega}_k | \bm{\theta}),\\
\end{equation}
where $K$ is the number of observed objects, and $\bm{\omega}_k$ is a vector of parameters that completely describe the $k$th planet's orbit. In our case, the two relevant parameters are $\msini$ and semi-major axis $a$, taken from the broader set that includes eccentricity, time of inferior conjunction, and argument of periastron. The Poisson likelihood can be understood as the product of the probability of detecting an observed set of objects (the product term in Equation 2) and the probability of observing no additional objects in the considered parameter space (the exponentiated integral). Equations 1 and 2 serve as the foundation for our occurrence model but do not take into account uncertainty in measurements of planetary orbits and minimum masses. In order to do this, we used \texttt{RadVel} and \texttt{emcee} to empirically sample the orbital posteriors of each system \citep{Fulton18, DFM13}. We hierarchically modeled the orbital posteriors of each planet in our catalog by summing our occurrence model over many posterior samples for each planet. The hierarchical Poisson likelihood is therefore approximated as

\begin{equation}
\mathcal{L}\approx e^{-\int \hat{\Gamma}(\bm{\omega} | \bm{\theta}) \,d\bm{\omega}} \prod_{k=1}^{K} \frac{1}{N_k} \sum_{n=1}^{N_k} \frac{\hat{\Gamma}(\bm{\omega}_k^n | \bm{\theta})}{p(\bm{\omega}_k^n | \bm{\alpha})},\\
\end{equation}
where $N_k$ is the number of posterior samples for the $k$th planet in our survey and $\bm{\omega}_k^n$ is the $n$th sample of the $k$th planet's posterior. $p(\bm{\omega} | \bm{\alpha})$ is our prior on the individual planet posteriors. We placed linear-uniform priors on $M$sin$i$ and log-uniform priors on $a$. We used \texttt{emcee} to sample our hierarchical Poisson likelihood.

We used two different occurrence frameworks to model our planet population. The first is a non-parametric model across bins uniformly spaced in ln($M$sin$i$) and ln($a$), with a set of steps $\bm{\Delta}$ of height $\bm{\theta}$. We define this framework with the occurrence function

\begin{equation}
\Gamma_N(\bm{\omega} | \bm{\theta}) = \theta_n | \bm{\omega} \in \Delta_n. \\ 
\end{equation}

The second framework is a broken power law as a function of semi-major axis, defined with the function

\begin{equation}
\Gamma_B(a | C, \beta, a_0, \gamma) = C (a/\mathrm{au})^{\beta}(1 - e^{-(a/a_0)^{\gamma}}), \\ 
\end{equation}
where $C$ is a normalization constant, $\beta$ is the occurrence power law index beyond the breaking point, $a_0$ determines the semi-major axis location of the breaking point, and $\beta + \gamma$ is the power law index within the breaking point. This model assumes a giant planet mass function that does not change with respect to semi-major axis. We fit this model to our population in order to explore whether giant planet occurrence falls off beyond the water-ice line.

\section{Results}
\label{sec:results}

\begin{figure*}[ht!]
\begin{center}
\includegraphics[width = 0.8\textwidth]{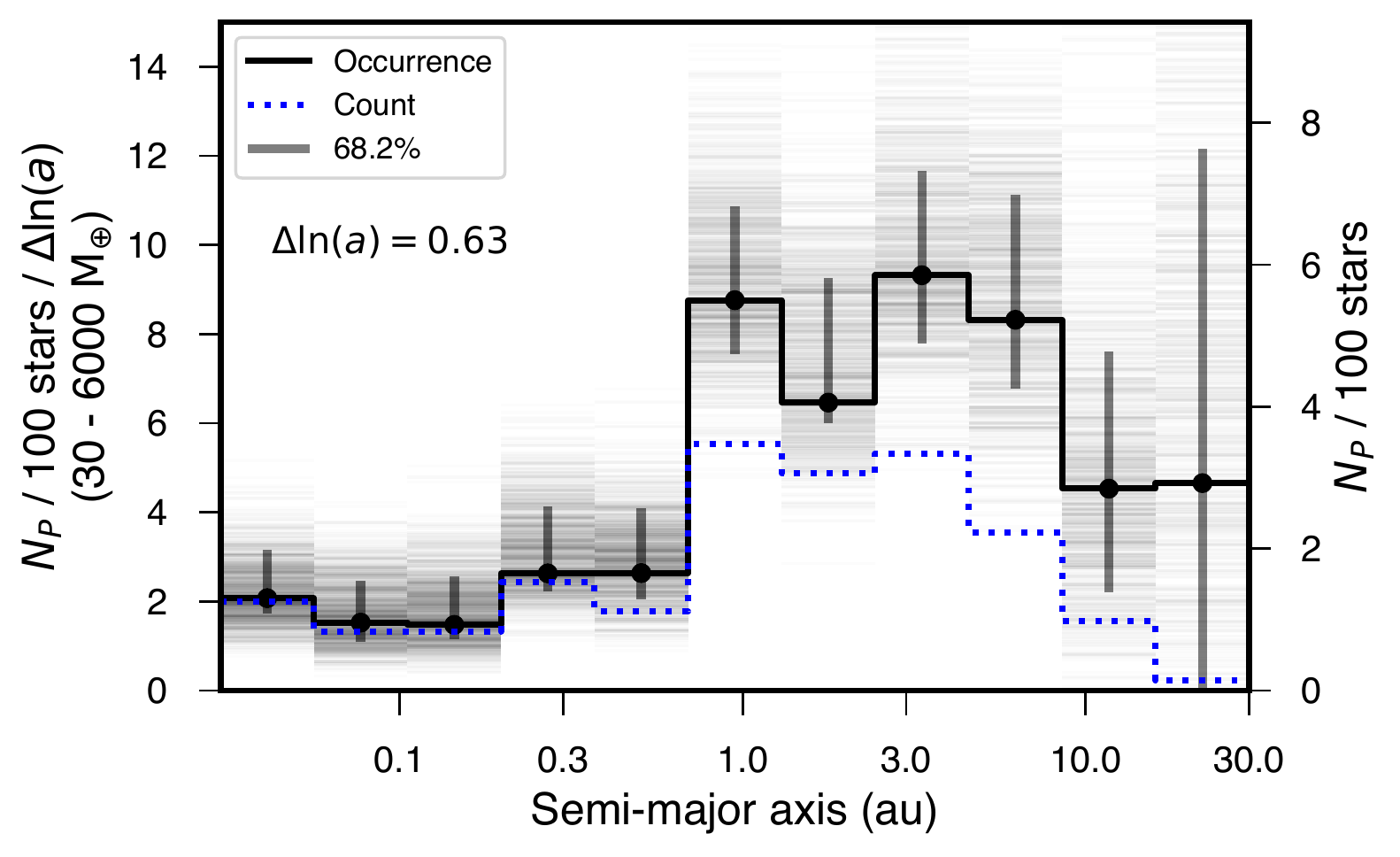}
\caption{Non-parametric occurrence rates for semi-major axes of 0.03--30 au for planets with minimum masses from  30--6000 \msini, assuming uniform occurrence across ln(\msini). The dashed blue line represents a planet count in each semi-major axis bin without correcting for completeness; bold lines and dots show the maximum posterior values for the Poisson likelihood model; vertical lines represent 15.9--84.1$\%$ confidence intervals (except for the last bin, which is not separated from zero and shows 0--68.2$\%$); and transparent steps show draws from the occurrence posterior. We see a clear enhancement around 1--10 au, and a tentative falloff beyond that range.}
\label{fig:semi_dist}
\end{center}
\end{figure*}

\begin{figure}[ht!]
\includegraphics[width = 0.49\textwidth]{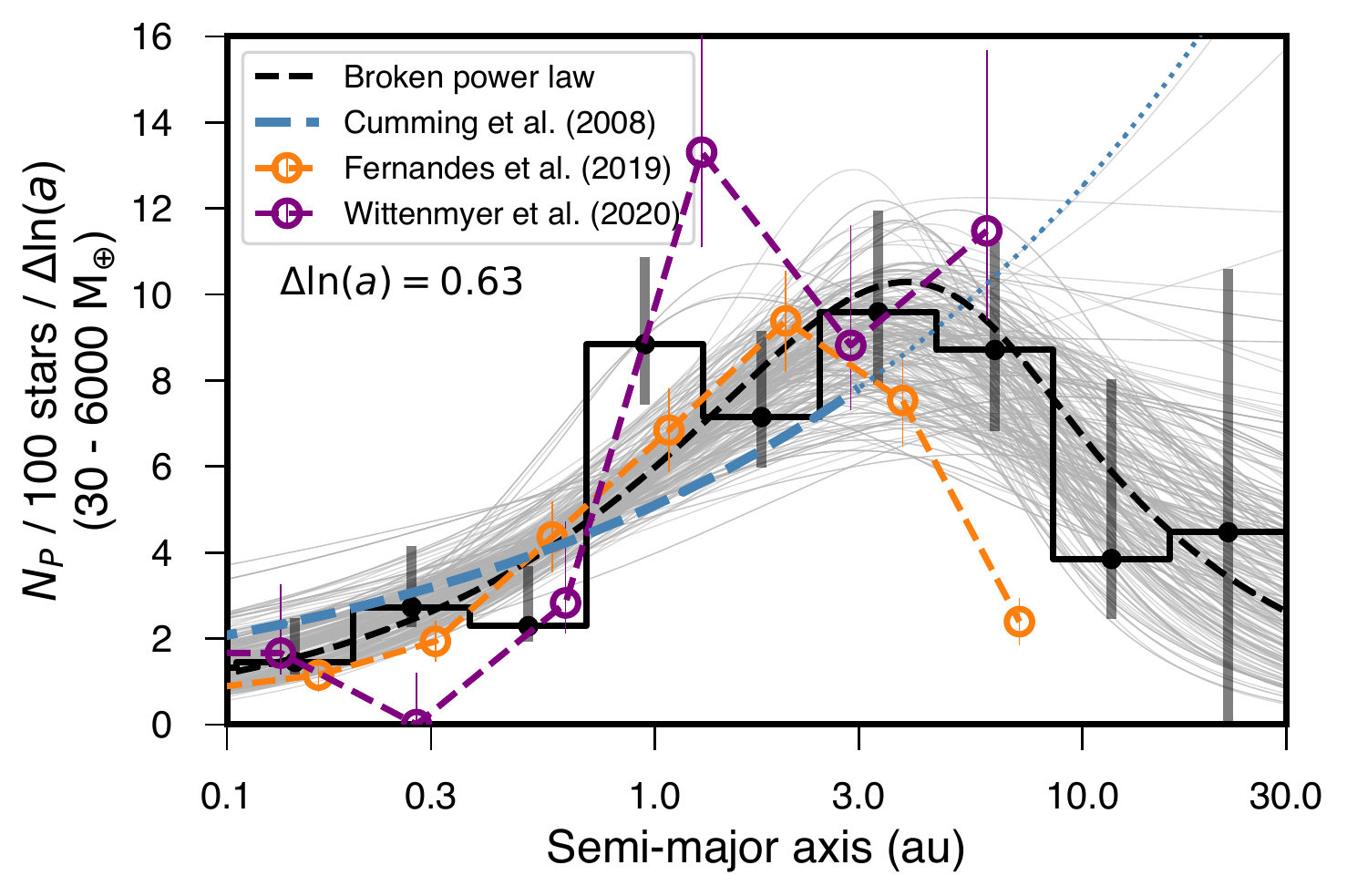}
\caption{Our broken power law model, juxtaposed with our non-parametric model and measurements from \cite{Fernandes19} and \cite{Wittenmyer20}. The transparent curves represent draws from the broken power law posterior. We find that the power law index beyond the break is $\sim$2.5$\sigma$-separated from zero, implying an occurrence falloff beyond the water-ice line. \citet{Cumming08} performed a power-law fit to the occurrence rates of planets orbiting only within 3 au; the light dotted blue line represents an extrapolation to wider separations.}
\label{fig:power_law}
\end{figure}

\begin{figure}[ht!]
\includegraphics[width = 0.5\textwidth]{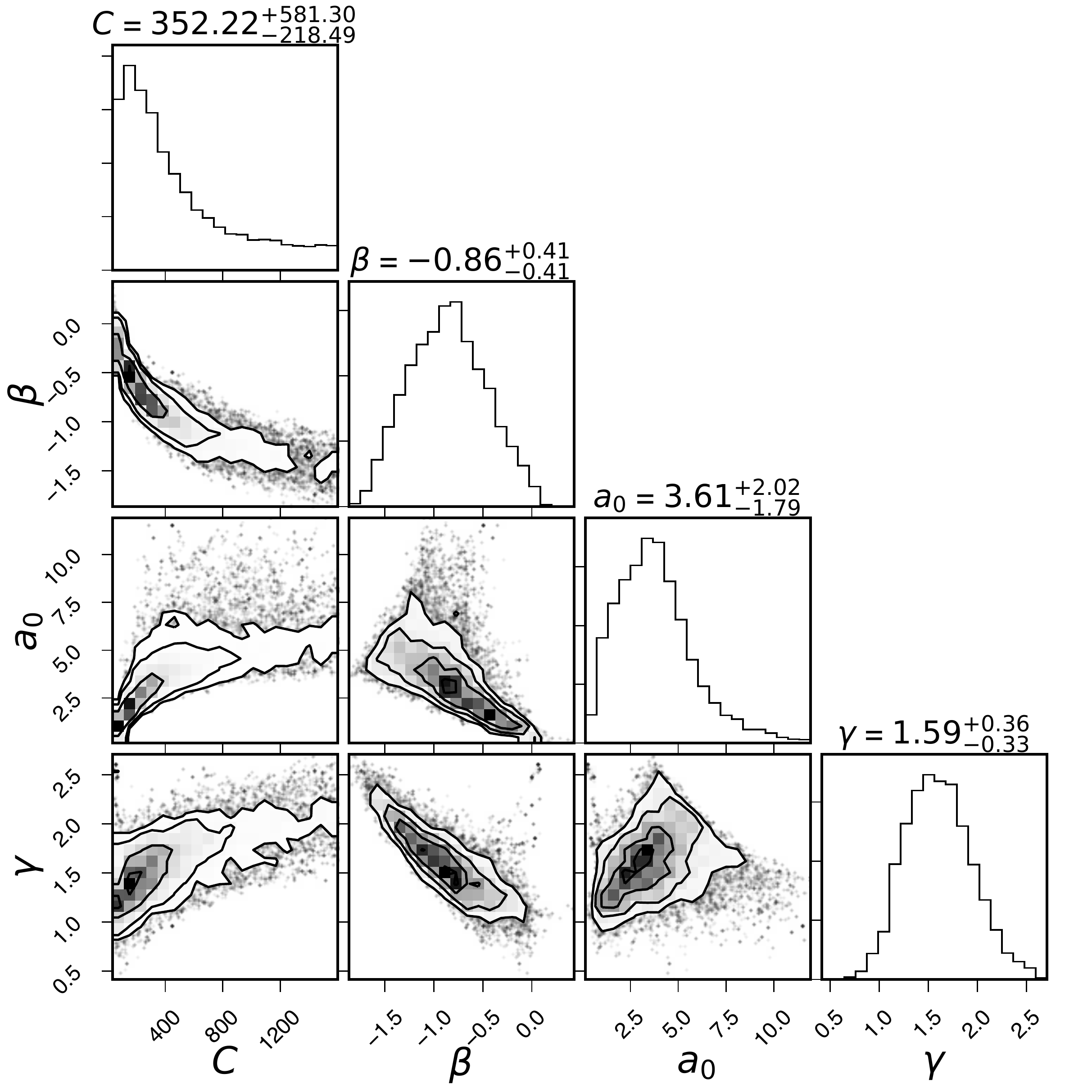}
\caption{Broken power law posterior. $C$ is a normalization constant, $\beta$ is the power law index beyond the break, $a_0$ determines the location of the break in units of au, and $\beta + \gamma$ is the power law index within the break. The index beyond the break $\beta$ is \deleted{$\sim2.5$ $\sigma$} \added{$\sim99.1\%$}-separated from zero.}
\label{fig:corner}
\end{figure}

\subsection{Enhancement for giant planets}
\label{sec:sub-jovians}

Figure \ref{fig:semi_dist} shows occurrence rates as a function of semi-major axis for planets with masses between 30 \mearthe\ and 6000 \mearthe, derived using the non-parametric model described in \S \ref{sec:occurrence} and assuming uniform occurrence across ln(\msini).  We confirmed the previous result from \citet{Wright09}, \cite{Cumming08}, \cite{Fernandes19}, and \cite{Wittenmyer20} that giant planet occurrence is enhanced by a factor of four beyond 1 au compared to within 1 au. \deleted{Planets more massive than Neptune} \added{Specifically, planets more massive than 30 \mearthe\ } are 2--4 times more common at orbital distances between 1--3 au relative to 0.1--0.3 au. \added{Using our broken power law model,} we find a median power law slope inside the break of 0.72$^{+0.16}_{-0.20}$, which is 2 $\sigma$ higher than the power law slope measured by \citet{Cumming08} (0.26$\pm$0.1). This difference is likely caused by the single power law model being pulled to lower values due to neglecting a flattening or turnover in occurrence at long orbital periods since \citet{Cumming08} was limited to planets orbiting inside 3 au.

\subsection{Distribution of giant planets beyond 3 au}
\label{sec:semi-dist}

Due to low completeness beyond our observational baselines, our occurrence results beyond 10 au are highly uncertain. However, we can estimate occurrence trends with the broken power law model described in \S \ref{sec:occurrence}. Figure \ref{fig:power_law} shows the broken power law results juxtaposed with the non-parametric results, and Figure \ref{fig:corner} presents the posteriors for the parametric model parameters. The medians and 68th percentile credible intervals for the broken power law model are listed in Table \ref{tab:broken}. Both assume uniform occurrence across ln(\msini). We find that 99.4\% of the posterior samples are consistent with a plateauing or declining occurrence rate beyond a peak around $3.6^{+2.0}_{-1.8}$ au. We find that the power law index beyond the peak is $\beta = -0.86^{+0.41}_{-0.41}$. This suggests a much shallower decline relative to the estimates of \cite{Fernandes19} but is also potentially discrepant with the constant prediction of \cite{Wittenmyer20}, as our model still measures a falloff. The results of our non-parametric fit are less clear, with integrated occurrence rates of $14.1^{+2.0}_{-1.8}$ and $8.9^{+3.0}_{-2.4}$ giant planets per 100 stars between 2--8 au and 8--32 au respectively. This suggests a fall-off in occurrence beyond 8 au with 1.5$\sigma$ confidence.

\begin{deluxetable}{lr}
\tabletypesize{\large}
\tablecaption{Broken Power-Law Model Parameters\label{tab:broken}}
\tablehead{
	\colhead{Parameter} & 
	\colhead{Value} 
}
\startdata
$C$ & $350^{+580}_{-220}$ \\
$\beta$ & $-0.86^{+0.41}_{-0.41}$ \\
$a_0$ & $3.6^{+2.0}_{-1.8}$ au \\
$\gamma$ & $1.59^{+0.36}_{-0.33}$ \\
\enddata
\end{deluxetable}

\subsection{Comparing sub- and super-Jovians}
\label{sec:mass-function}

\begin{figure}[ht!]
\includegraphics[width = 0.49\textwidth]{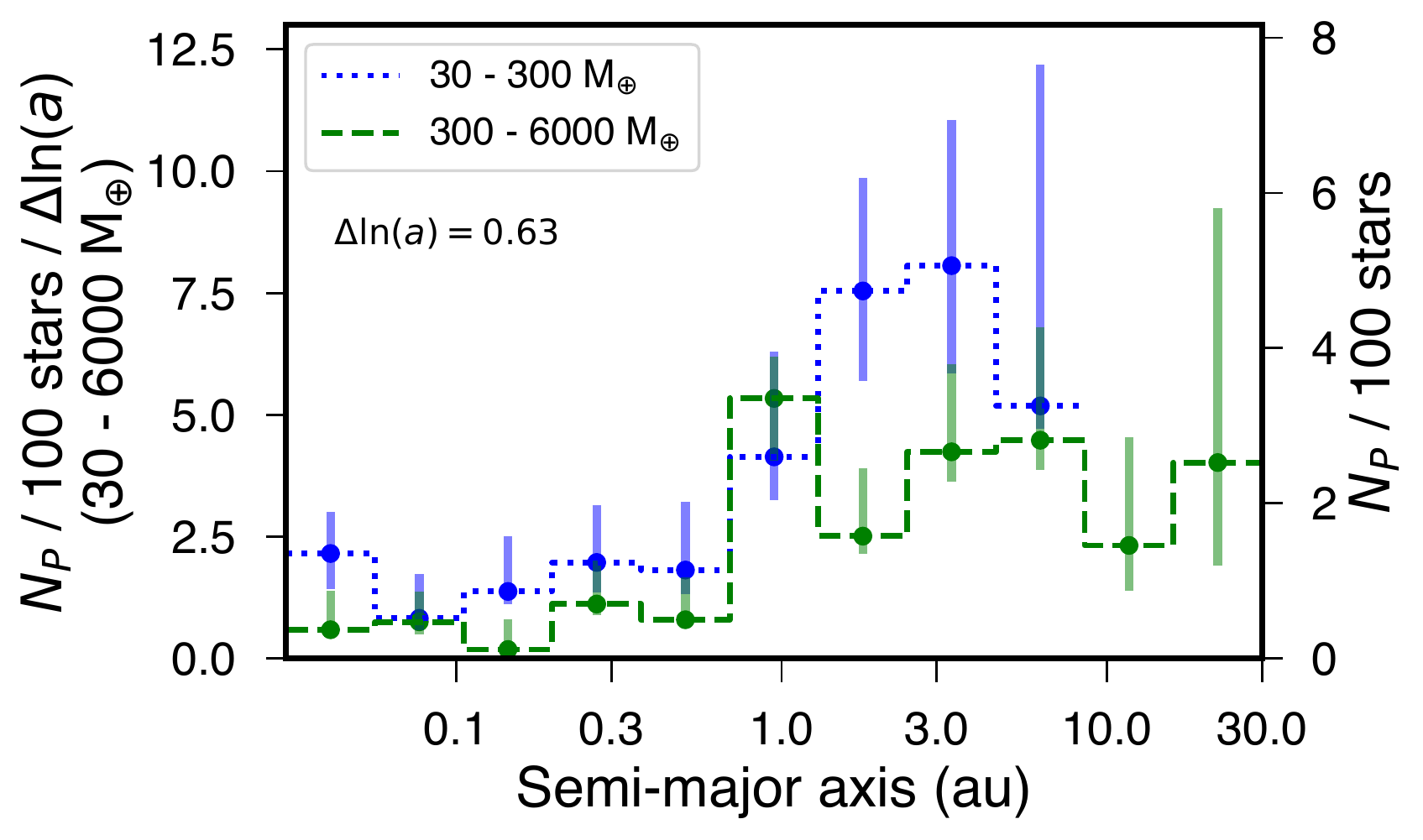}
\caption{A comparison between sub- and super-Jovian occurrence. Steps and dots show maximum posterior values, and vertical lines show 15.9--84.1$\%$ confidence intervals. The sub-Jovians are consistently more common than the super-Jovians, and both populations are enhanced beyond 1 au. Combining these two populations produces the same trends seen when we assume uniform occurrence across all masses.}
\label{fig:semi_dist_sub}
\end{figure}

Figure \ref{fig:semi_dist_sub} compares non-parametric occurrence rates for giant planets more and less massive than 300 \mearthe. We find a quantitatively similar occurrence enhancement around 1--10 au for both the sub-Jovian-mass and Jovian-mass planets. However, we lack the sensitivity to measure the occurrence rate of  sub-Jovian mass planets beyond 10 au, to assess whether they exhibit the fall-off in occurrence at large orbital separations seen when examining occurrence across both mass ranges. The sub-Jovian planets are more common than the super-Jovian planets across a wide range of separations, particularly beyond the water ice line. We find a similar enhancement for sub-Saturns below 150 \mearthe, implying that this occurrence enhancement is independent of planet mass.

\begin{figure}[ht!]
\includegraphics[width = 0.49\textwidth]{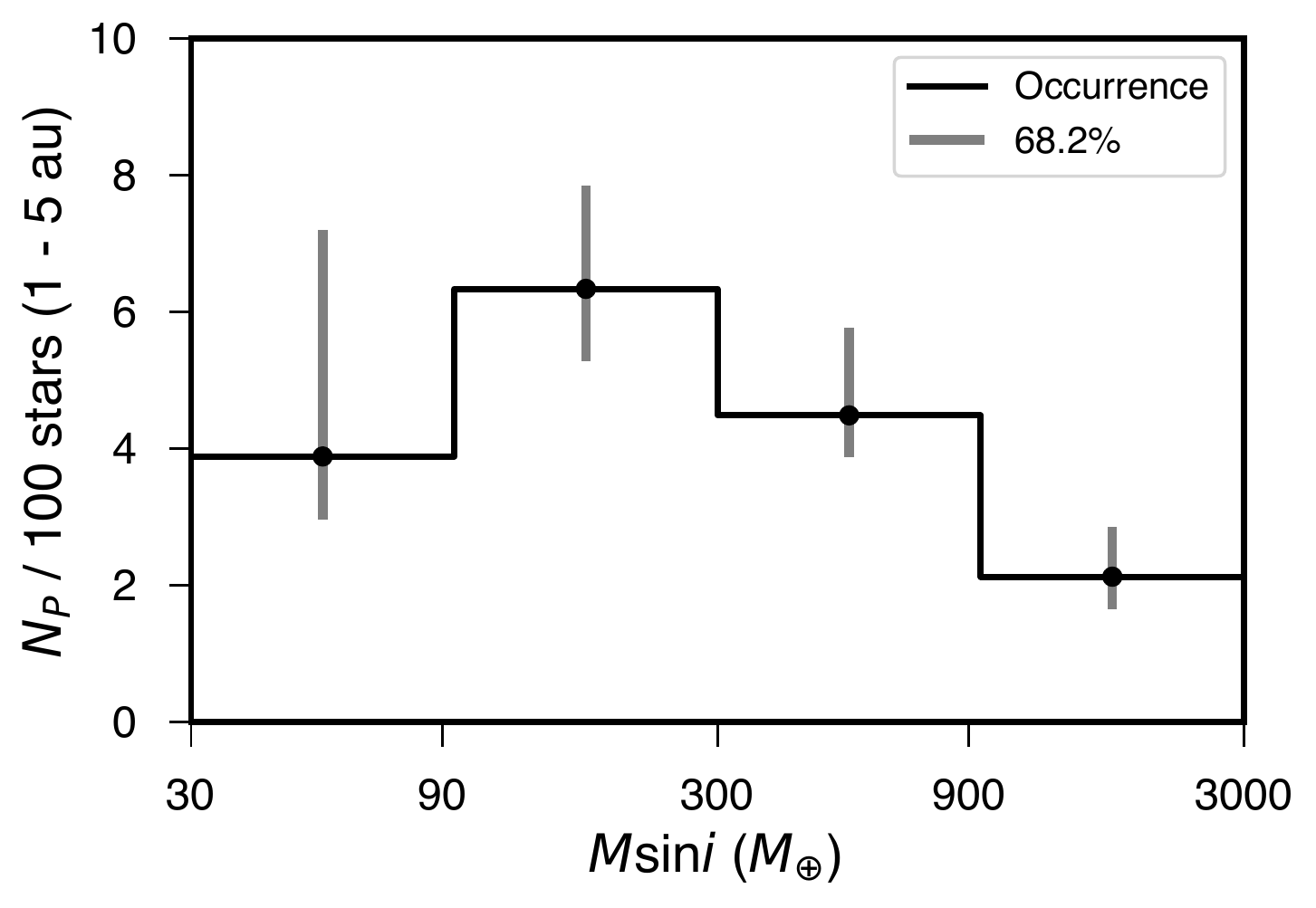}
\caption{Planet occurrence within 1--5 au with respect to \msini . Steps and dots show maximum posterior values, and vertical lines show 15.9--84.1$\%$ confidence intervals. The mass function is constant within 30--900 \mearthe, and falls off beyond 900 \mearthe .}
\label{fig:beyond-ice-mass-function}
\end{figure}

We more concretely measured occurrence as a function of mass by performing a non-parametric fit to our sample within 1--5 au. Figure \ref{fig:beyond-ice-mass-function} shows occurrence as a function of \msini\ within 30--3000 \mearthe, in four steps. This figure shows that our assumption of a uniform ln(\msini) distribution beyond the ice line is valid up to 900 \mearthe, but the distribution falls off with $\sim$2$\sigma$ significance above 900 \mearthe. If this is also true beyond 5 au, where low completeness prevents us from making a similar measurement, then we may be underestimating broad giant planet occurrence in our lowest-completeness region of parameter space, beyond 10 au. This is because our only detections in that regime are more massive than 300 \mearthe, and all but one of them are more massive than 900 \mearthe. 

\subsection{Occurrence with respect to stellar mass and metallicity}
\label{sec:mass-and-metallicity}

\begin{figure}[ht!]
\includegraphics[width = 0.49\textwidth]{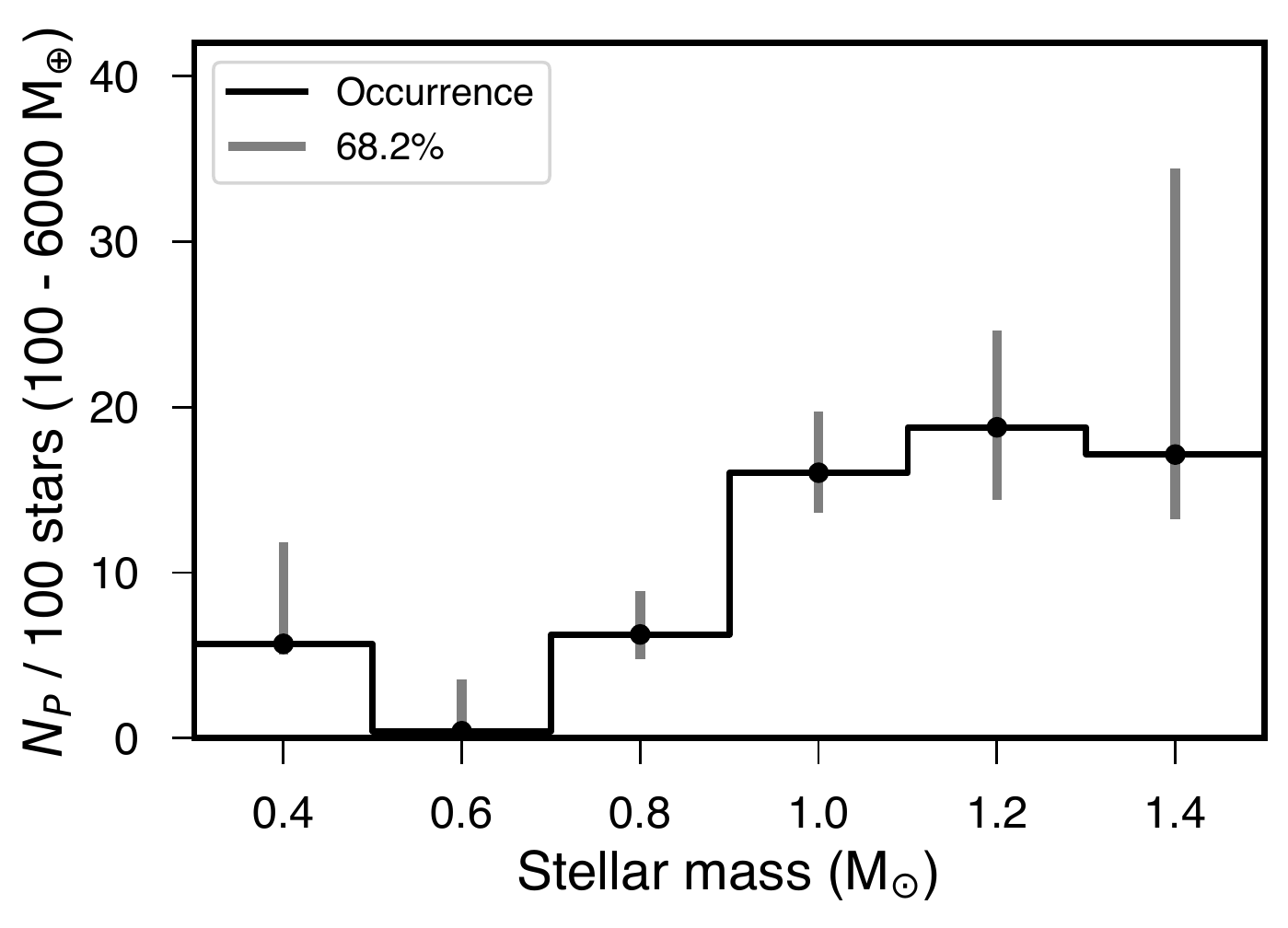}
\caption{Occurrence of giant planets more massive than 100 \mearthe\ and within 1--5 au as a function of host star mass, in six splits. Steps and dots show maximum posterior values, and vertical lines show 15.9--84.1$\%$ confidence intervals. There is an increase in occurrence beyond roughly 1 \Msun, which is in agreement with \cite{Johnson10}'s original measurement of giant planet occurrence versus host-star mass.}
\label{fig:giants_mass}
\end{figure}

\begin{figure}[ht!]
\includegraphics[width = 0.49\textwidth]{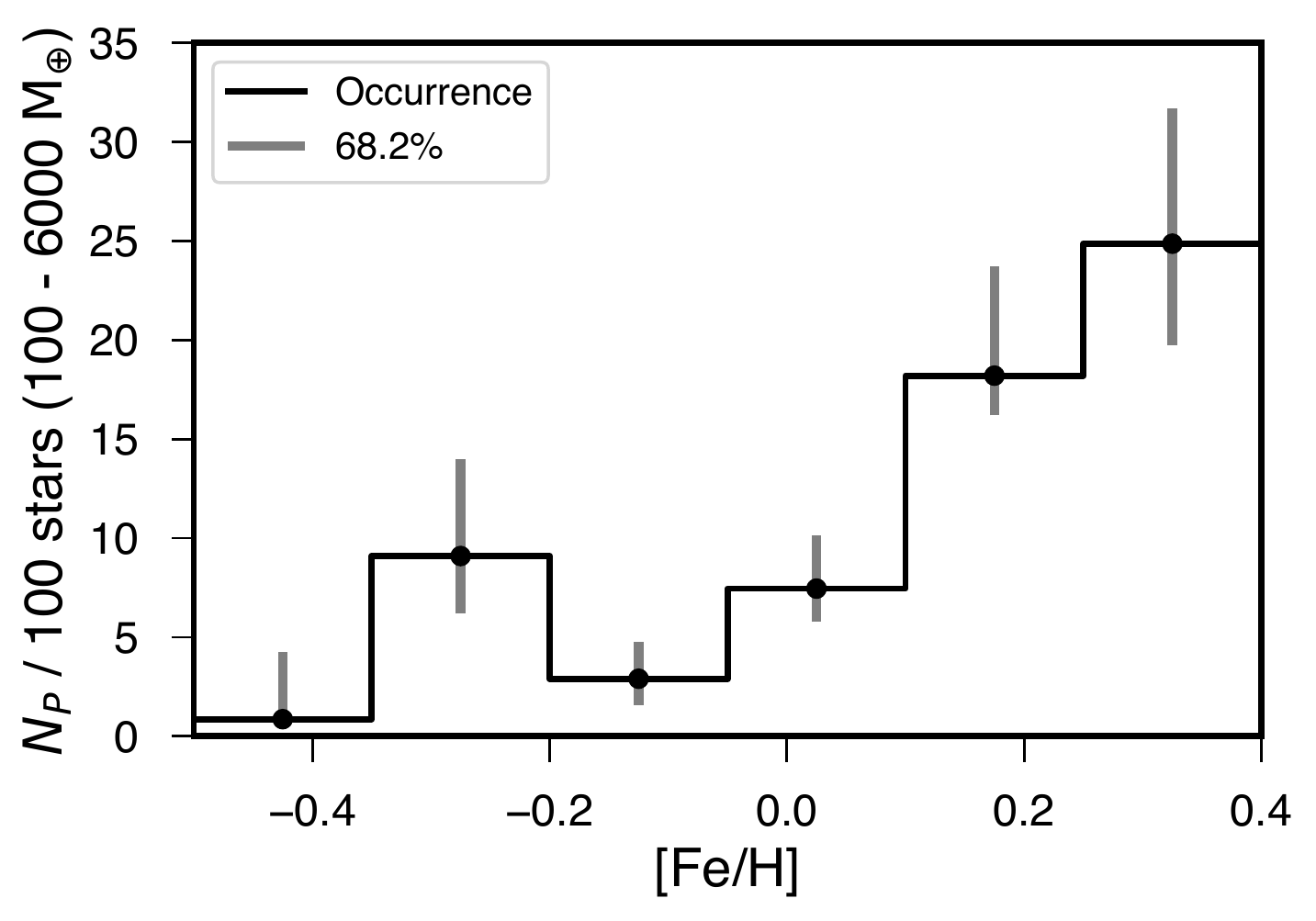}
\caption{Occurrence of giant planets more massive than 100 \mearthe\ and within 1--5 au as a function of host star metallicity, in six splits. Steps and dots show maximum posterior values, and vertical lines show 15.9--84.1$\%$ confidence intervals. There is a clear increase in occurrence beyond roughly 0.1 dex, which is in agreement with \cite{Fischer05}'s original report of a correlation between giant planet occurrence and host-star metallicity.}
\label{fig:giants_metallicity}
\end{figure}

In addition to measuring occurrence with respect to semi-major axis and \msini, we measured the broad occurrence rate of giant planets more massive than 100 \mearthe\ and within 1--5 au with respect to host-star mass and metallicity. \added{We chose a lower limit of 100 \mearthe\ instead of 30 \mearthe\ in order to restrict our analysis to search-complete regions within 1--5 au, since 30 \mearthe\ planets are effectively undetectable beyond 3 au.} For each of these two stellar properties, we computed occurrence across six divisions, in steps of 0.2 \Msun across 0.3--1.5 \Msun and 0.15 dex across -0.5--0.4 dex respectively. Figure \ref{fig:giants_mass} shows occurrence with respect to host-star mass, while Figure \ref{fig:giants_metallicity} shows occurrence with respect to host-star [Fe/H]. Both of our measurements agree with prior results. \cite{Johnson10}\added{, whose stellar sample was excluded from CLS due to its bias toward giant planet hosts,} measured giant planet occurrence across stellar mass and found an increase in occurrence with increasing stellar mass beginning near 1 \Msun. \added{\cite{Wittenmyer20b} independently found an increase in giant planet occurrence beyond 1 \Msun.} We see the same phenomenon in our sample, as presented in Figure \ref{fig:giants_mass}. Similarly, \cite{Fischer05} found that giant planet occurrence increases with increasing [Fe/H] beyond 0.1 dex, \added{as did \cite{Reffert15} and \cite{Wittenmyer16}.} We see the same transition near 0.1 dex in Figure \ref{fig:giants_metallicity}. 

\section{Discussion}
\label{sec:discussion}

\subsection{Comparison to previous RV surveys}
\label{sec:rv-surveys}
The last few years have seen a number of RV studies examining the population of long-period planets. \cite{Fernandes19} probed planet occurrence as a function of orbital period by extracting planetary minimum masses and periods, as well as completeness contours, from a catalog plot shown in \cite{Mayor11}, which presented a HARPS \citep{Mayor03} and CORALIE \citep{Baranne96} blind radial velocity survey of 822 stars and 155 planets over 10 years (corresponding to a 4.6 au circular orbit around a Solar-mass star). \cite{Mayor11}, who did not publish their HARPS and CORALIE RVs, measured giant planet occurrence as a function of orbital period out to 4000 days, in the range of the water-ice line. \cite{Fernandes19} pushed out to low-completeness regimes and estimated a sharp falloff in occurrence beyond the water-ice line. They measured an integrated occurrence rate of 1.44$\pm$0.54 giant planets (0.1--20 $M_J$) per 100 stars for separations between 3.8 and 7.1 au. Our results indicate a much higher occurrence rate for the same planets at those separations; 15.5$^{+3.2}_{-3.0}$ giant planets per 100 stars. The treatment of partial orbits in \cite{Mayor11} is unclear, and they only measured occurrence with respect to orbital period out to 3000 days ($\sim$4 au). If \cite{Mayor11} under-reported partial orbits beyond this period in their sample or overestimated sensitivity to partial orbits, then that could explain the large discrepancy between this work and \citet{Fernandes19} at separations beyond 10 au.

In contrast, \cite{Wittenmyer20}, which drew from the Anglo-Australian Planet Search \citep{Tinney01} to construct a blind survey of 203 stars and 38 giant planets over 18 years, found that giant planet occurrence is roughly constant beyond the water-ice line, out to almost 10 au. \cite{Wittenmyer20} reports an occurrence rate of $6.9^{+4.2}_{-2.1}$ giant planets \deleted{(0.1--20 $M_J$)} \added{> 0.3 $M_J$} per 100 stars with periods between 3000 and 10,000 days ($\approx$4--9 au). Our integrated occurrence rate in the same region of parameter space is slightly higher at $12.6^{+2.6}_{-2.0}$ giant planets per 100 stars but it is consistent to within 1 $\sigma$ with the \cite{Wittenmyer20} result.

\subsection{Comparison to Kepler survey}
\label{sec:kepler}
\cite{Foreman-Mackey16} performed an automated search for long-period transiting exoplanets in a set of archival \textit{Kepler} light curves of G and K stars. For planets between 1.5--9 au and 0.01--20 $M_J$ and using a probabilistic mass-radius relationship drawn from \cite{ChenK16}, they found an occurrence rate density of $\frac{\mathrm{d}^2N}{\mathrm{dln}(a)\mathrm{dln}(M)} = 0.068 \pm 0.019$. We applied our occurrence model to the same parameter space and found $\frac{\mathrm{d}^2N}{\mathrm{dln}(a)\mathrm{dln}(\msini)} = 0.0173 \pm 0.0022$. The \textit{Kepler} measurement is $2.66\sigma$ separated from our measurement. We are far less sensitive to planets in the 0.01--0.1 $M_J$ regime than \cite{Foreman-Mackey16}; this may partly explain the discrepancy in our results.

\subsection{Comparison to direct imaging surveys}
\label{sec:direct-imaging}
RV surveys have recently begun to approach baselines long enough to detect and place limits on the frequency of planets like those detected by direct imaging. One caveat is that direct imaging surveys usually target stars younger than 100 Myr, while RV surveys generally target stars older than 1 Gyr. Young planets retain significant heat from their formation and are bright in the infrared wavelengths covered by direct imaging surveys. However, young stars also tend to be active and rapidly rotating, which makes precise RV work difficult. Because of this, there is minimal overlap between planets that have been detected by direct imaging and planets that have been detected by radial velocity measurements.

\begin{figure*}[ht!]
\includegraphics[width = 0.49\textwidth]{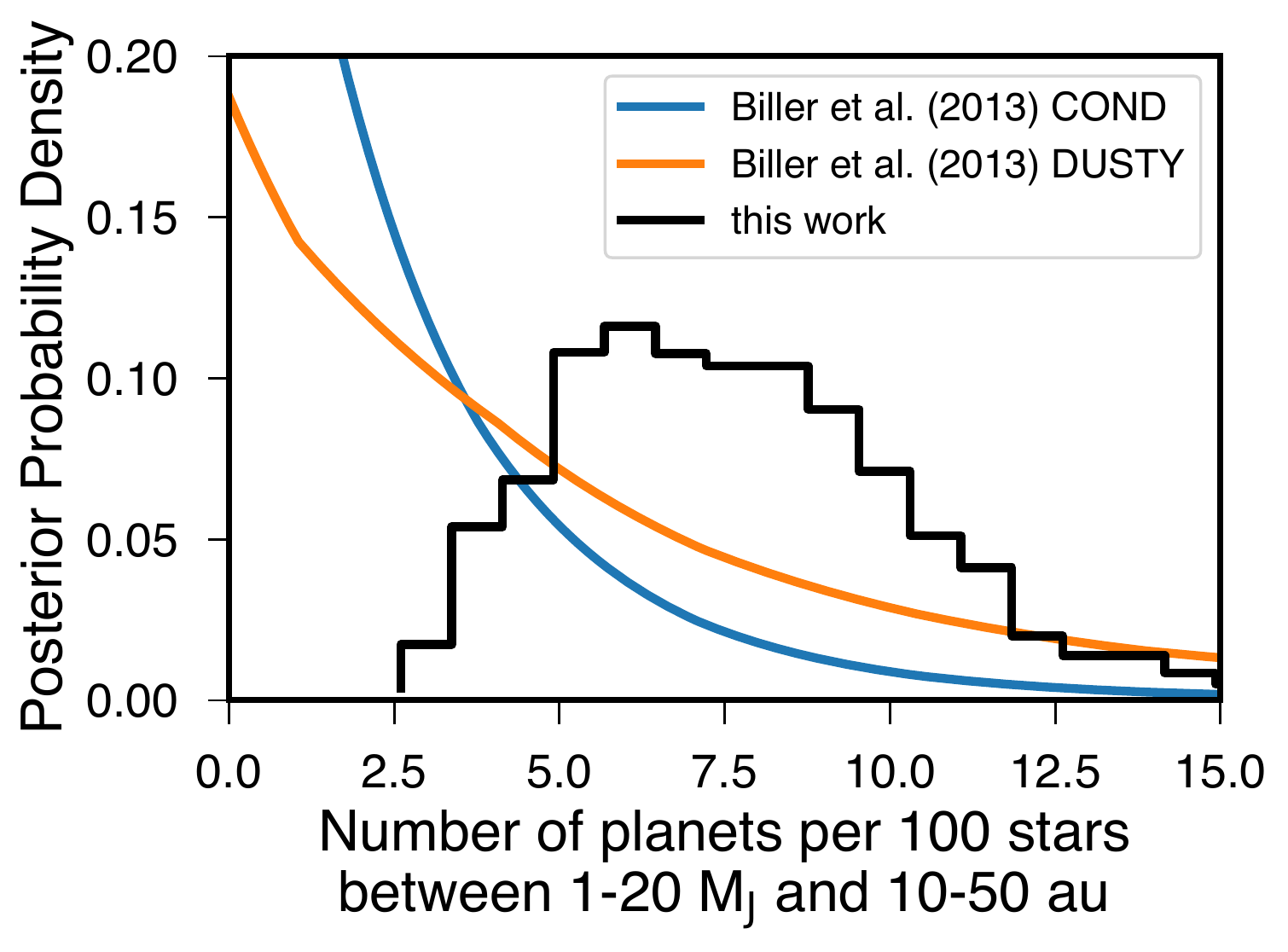}
\includegraphics[width = 0.49\textwidth]{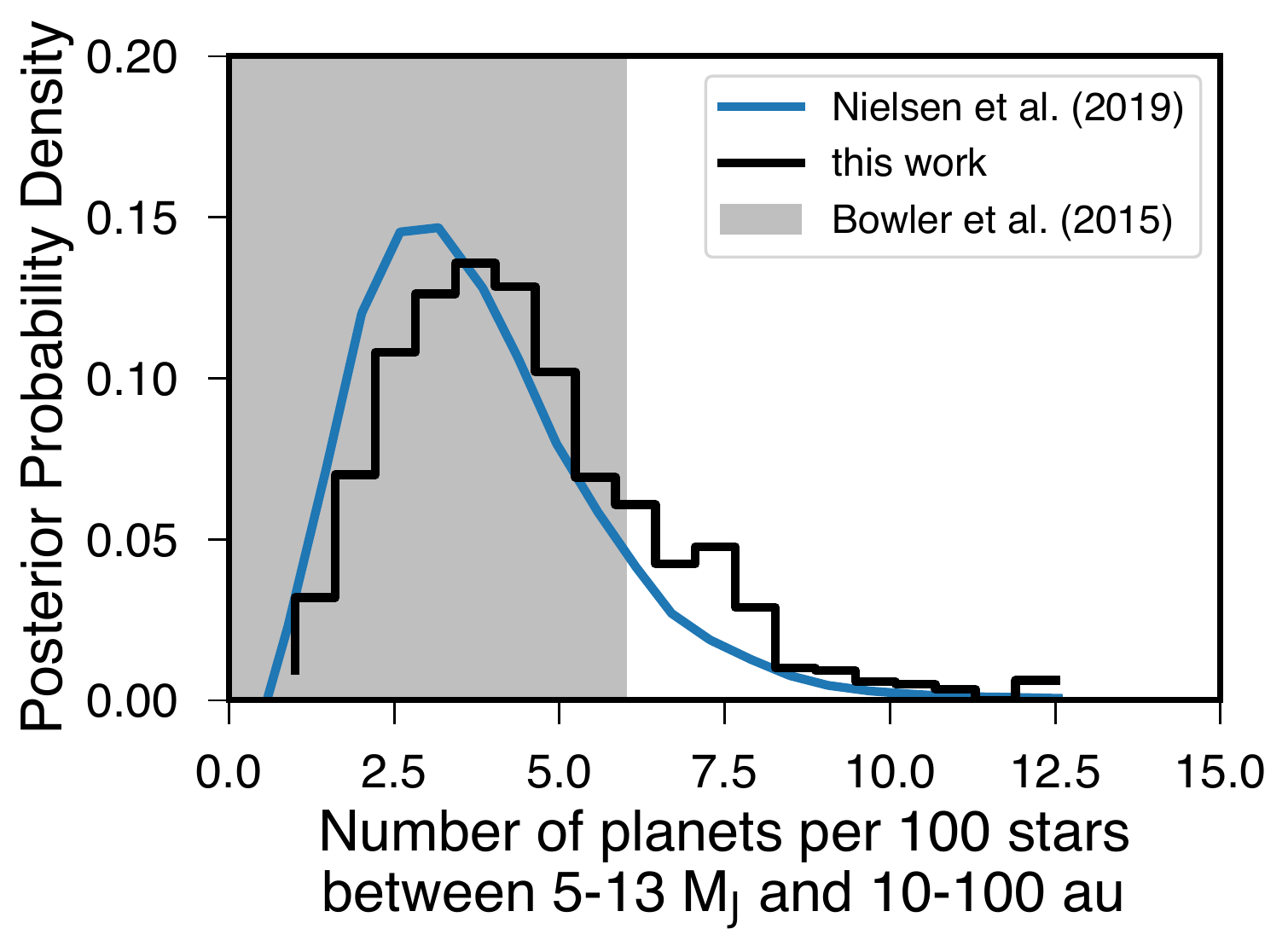}
\caption{Occurrence rate comparison to direct imaging studies. \emph{Left}: Frequency of cool, massive companions with the direct imaging study of \citet{Biller13}. While they did not detect any planets in their survey they were able to put upper limits on the frequency of companions using assumptions of either hot-start (COND) or cold-start (DUSTY) models for planetary formation and infrared brightness.
\emph{Right}: Same as left, but compared with the results of \citet{Bowler15} and \citet{Nielsen19} for the mass and separation limits specified in the x-axis label. The gray shading represents the 95$\%$ upper limit on occurrence from \citet{Bowler15}.}
\label{fig:di_comp}
\end{figure*}

We can still compare rates across these detection methods by making the assumption that giant planet occurrence does not change as host stars age beyond $\sim$10 Myr, once protoplanetary disks have dissipated. We compared our occurrence model to the results of two direct imaging surveys of nearby stars. \cite{Biller13} imaged 80 stars in nearby moving groups and detected a small number of brown dwarf companions but no planetary-mass companions. They used stellar evolution and planet formation models to estimate constraints on cold giant occurrence from their nondetections and sensitivity. More recently, \cite{Nielsen19} imaged 300 stars and detected six planets and three brown dwarfs. Figure \ref{fig:di_comp} compares these results to our occurrence measurements in their respective regions of parameter space. Our measurements are compatible with the limits placed on planets with masses 1--20 $M_J$ and separations between 10--50 au by \cite{Biller13}, depending on their assumed stellar evolutionary model that determines the expected brightness of young giant planets. Our measurement for planets with masses 5--14 $M_J$ orbiting between 10--100 au is in excellent agreement with the results of \cite{Nielsen19}. The only shared quality of our modeling methods is a Poisson counting likelihood. With the caveat of small number statistics, this is a remarkable benchmark for comparing exoplanet occurrence across independent search methods.

\begin{figure*}[ht!]
\includegraphics[width = 0.49\textwidth]{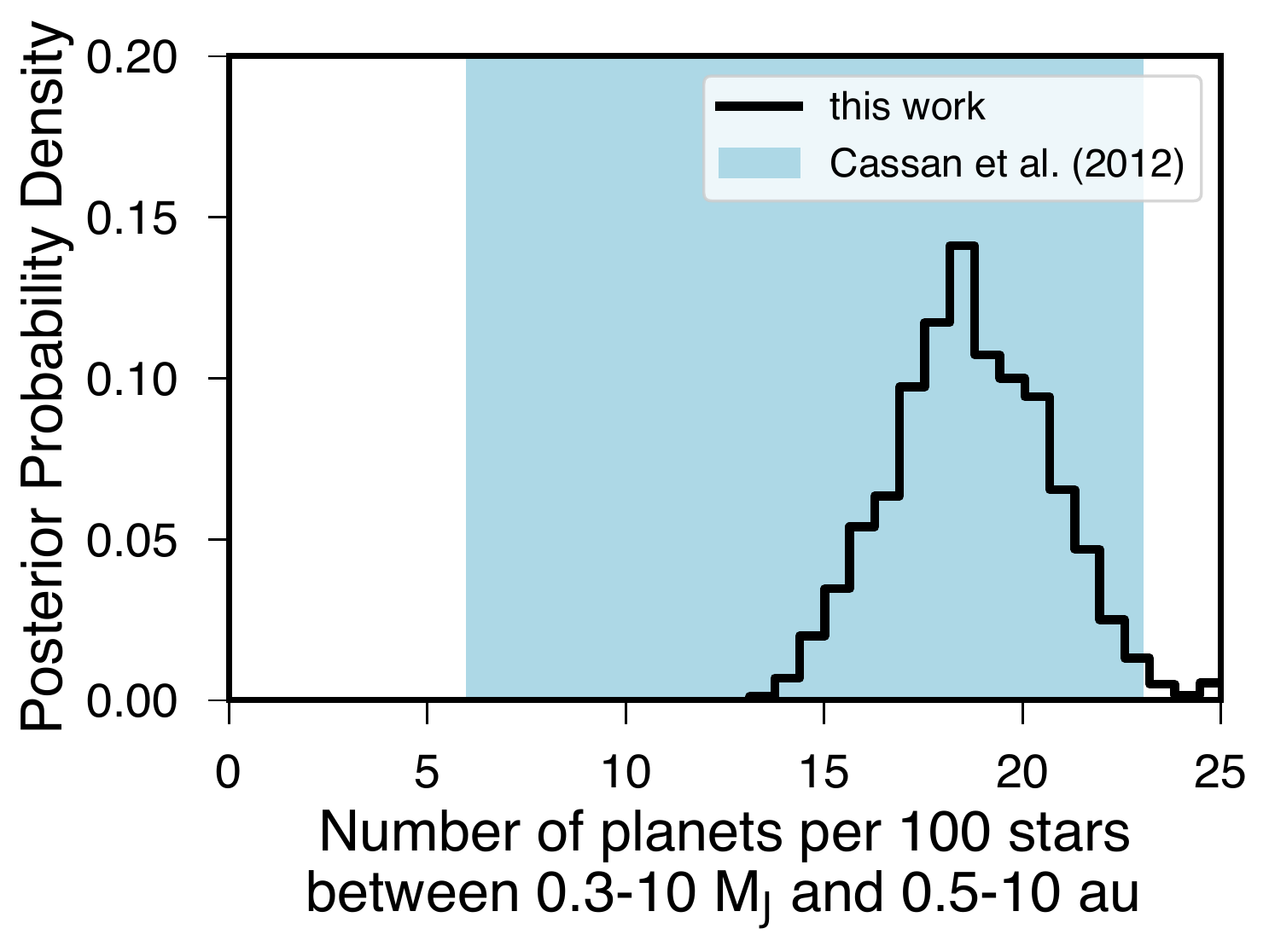}
\includegraphics[width = 0.49\textwidth]{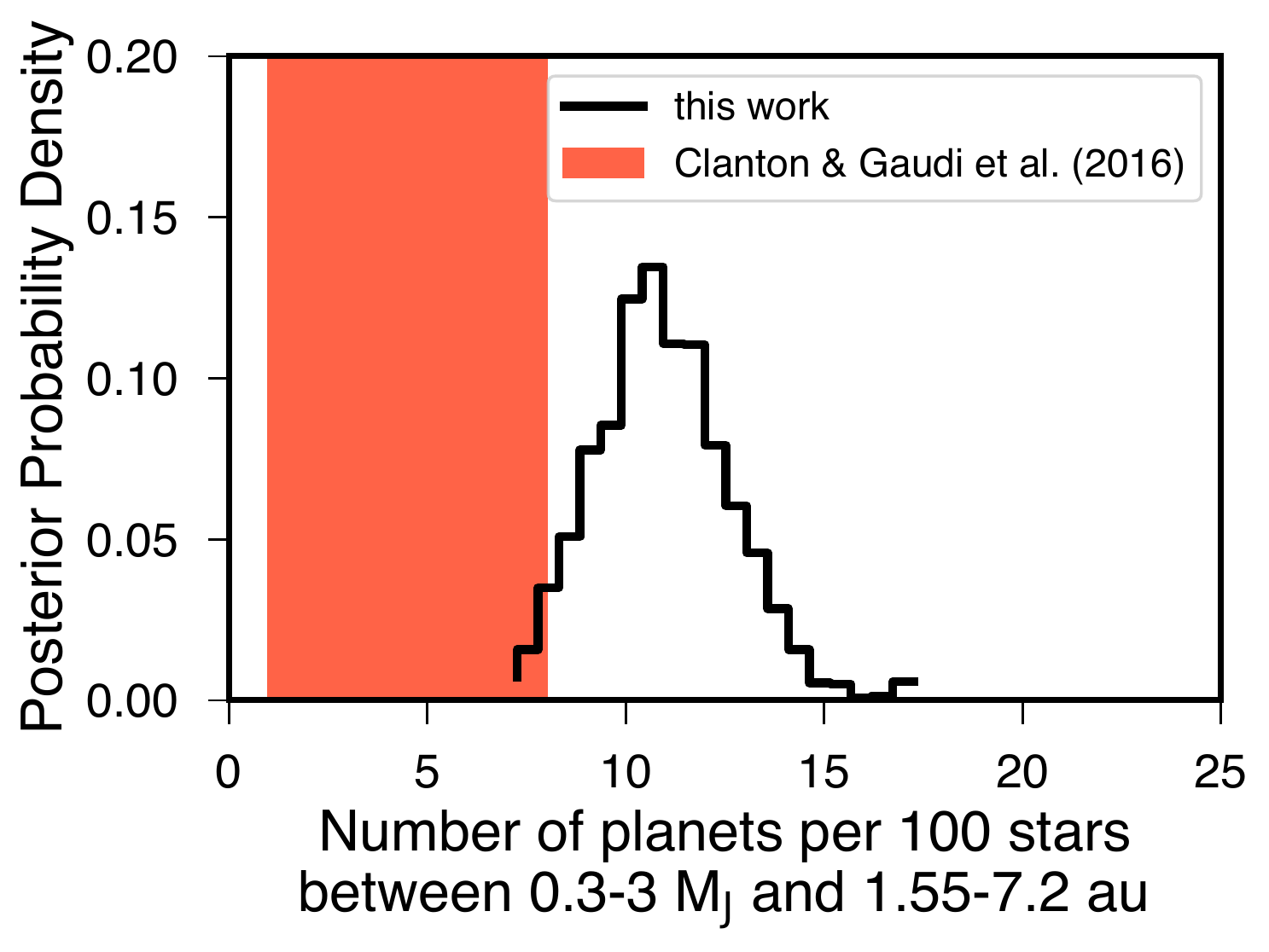}
\caption{\emph{Left:} Occurrence rate comparison with the microlensing survey of \citet{Cassan12}. We plot the 1 $\sigma$ limits from \citet{Cassan12} as the shaded blue region. The occurrence rate posterior from this work is plotted in black.
\emph{Right:} Occurrence rate comparison with the combined analysis of \citet{Clanton16}. The occurrence rate posterior from this work is plotted in black. The 1 $\sigma$ limits from \citet{Clanton16} are indicated by the shaded red region. \citet{Clanton16} combine constraints from direct imaging, microlensing, and previous radial velocity studies.}
\label{fig:ml_comp}
\end{figure*}

\subsection{Comparison to gravitational microlensing surveys}
\label{sec:microlensing}

We compare our model to the microlensing surveys of \cite{Cassan12} and \cite{Clanton16}. Like all gravitational lensing surveys, these studies assume a broad prior for stellar type based on Galactic observations, a prior that peaks in the M dwarf range. Our planet-hosting stars have a much higher median mass than this range, but since the gravitational lensing estimates comes purely from a galactic model prior, we chose to perform this broad comparison across stellar masses with the knowledge that the mass range for the lensing numbers is poorly constrained. Figure \ref{fig:ml_comp} shows that our estimates agree with broad constraints from the pure lensing survey \citep{Cassan12}. On the other hand, the constraints of \cite{Clanton16} strongly disagree with our planet occurrence measurement in the same parameter box. This may be due to that study having a significantly better constrained sample of M dwarfs, which would separate their stellar sample from our broader FGKM sample. deleted{\cite{Montet14} performed an RV survey of M dwarfs} \added{\cite{Endl06}, \cite{Bonfils13}, and \cite{Montet14} performed independent RV surveys of M dwarfs} and all showed that M dwarfs have a significantly lower giant planet occurrence rate than more massive stars. This implies that a survey of M dwarfs should yield a lower giant planet occurrence rate than a broad survey of FGKM stars, and this is exactly what we see in our comparison to \cite{Cassan12}.

\subsection{Implications for planet formation}
\label{sec:formation}

\citet{Cumming08} first identified an enhancement in the occurrence rate of giant planets beyond orbital periods of $\sim$300 days. We expect such enhancements based on planetary migration models \citep{Ida04a}. The orbital period distribution in \cite{Cumming08} predicted a smooth rise in occurrence toward longer orbital periods, but we observed a sharp transition around 1 au, as seen in Figure \ref{fig:semi_dist}. \citet{Ida_Lin08_v} later suggested that additional solid materials due to ices in the protoplanetary disk could augment the formation of gas giant planets and cause a rapid rise in the occurrence rate of these planets beyond the water-ice line. 

\begin{figure}[ht!]
\includegraphics[width = 0.5\textwidth]{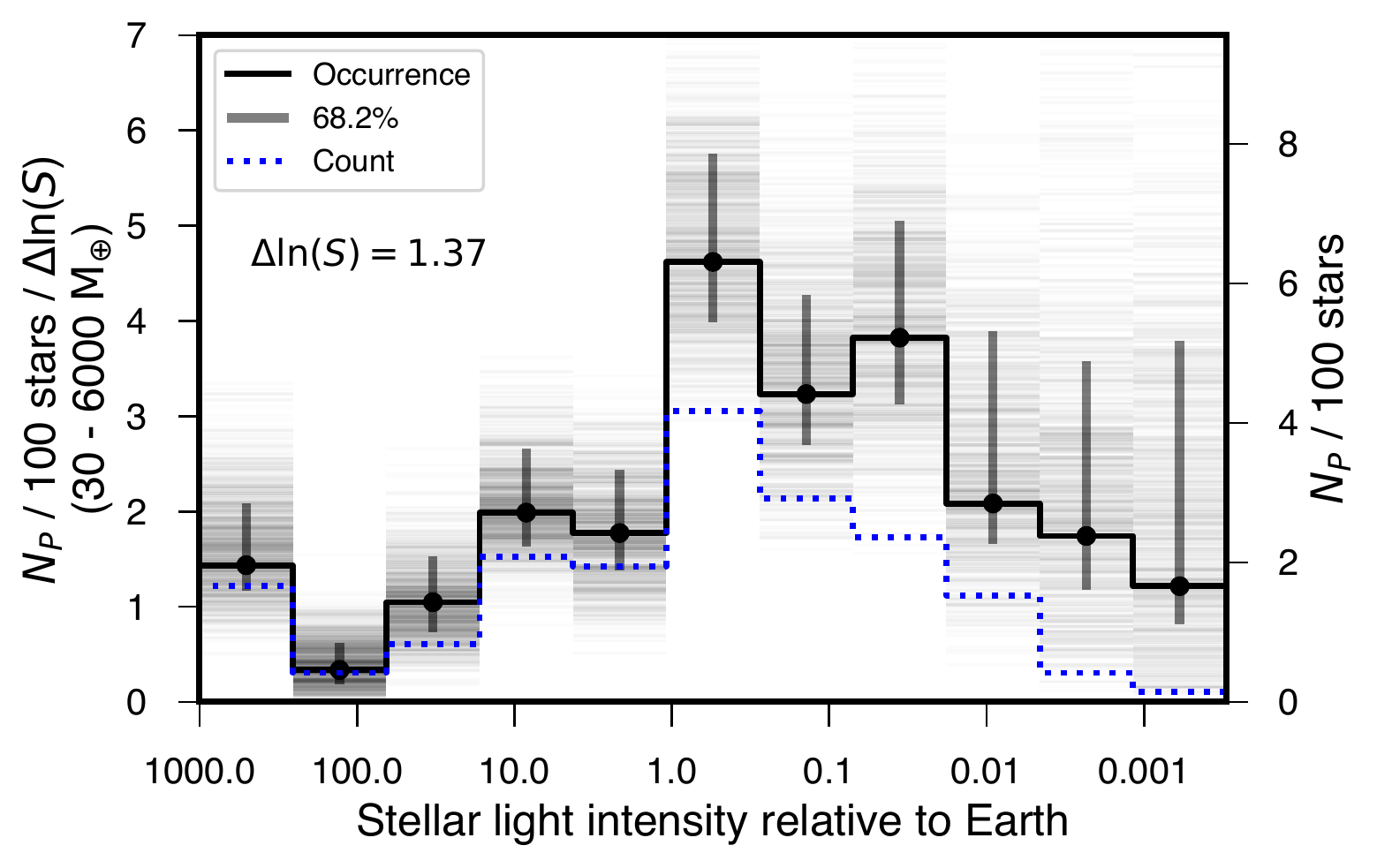}
\caption{Analogous to Figure 2, occurrence with respect to stellar light intensity instead of orbital separation. Here we see a similar enhancement in the occurrence rate of giant planets where the insolation flux is equal to that of Earth and tentative evidence for a fall off in occurrence just beyond that.}
\label{fig:insol_hist}
\end{figure}

If increased solids near and beyond the ice line cause a sharp rise in the occurrence rate, then we might expect this rise to be more well-defined when looking in a unit more closely related to the temperature in the protoplanetary disk. In Figure \ref{fig:insol_hist}, we plot the occurrence rate as a function of stellar light intensity relative to Earth. The occurrence rate with respect to flux is qualitatively similar to the rate with respect to orbital separation. We do not see strong evidence that the occurrence rate enhancement is any more localized in terms of stellar light intensity relative to Earth. The decline in occurrence for fluxes less than that received by the Earth from the Sun looks more convincing, but all except for the one bin with the highest occurrence near 1 au are consistent with a constant occurrence rate.

We can separate the puzzle of gas giant formation into two components: the growth of solid cores that are large enough to undergo runaway gas accretion, and the process of gas accretion onto solid cores. It is currently unclear whether giant planet occurrence increases beyond the ice line because cores form more easily in this region, or because conditions are more favorable for rapid gas accretion onto solid cores. A number of studies \citep[e.g.][]{Morbidelli15,Schoonenberg17,Drazkowska17} have argued that that core growth should be enhanced beyond the ice line.  If the solid grain sizes and densities beyond the ice line are enhanced during the earliest stages of planet formation, it would facilitate pebble clumping that leads to planetesimal formation and also result in higher pebble accretion rates onto the growing core \citep[e.g.][]{Bitsch19}. 

It is also possible that gas giants are more common beyond the ice line because it is easier for cores to rapidly grow their gas envelopes in this region. The rate at which the growing planet's envelope can cool and contract (hence accreting more gas) depends sensitively on its envelope opacity \citep[e.g.][]{Bitsch21}.  In a recent study, \cite{Chachan21} used dust evolution models to study the effect of dust opacity and dust-to-gas ratio on giant planet formation in the epoch immediately following the end of core formation. They found that as the disk evolves, decreasing dust opacity beyond the water-ice line allows for higher gas accretion rates in this region. 

\citet{Ida18} recently updated their models with an improved treatment of Type II migration. This mechanism would produce a broad semi-major axis distribution with many giant planets migrating inward to separations less than 1 au. However, \citet{Fernandes19} show that this model does not agree well with the occurrence contrast between the peak and the inner regions of these systems. Our results are in close agreement with those of \citet{Fernandes19} for separations less than 3 au. The multi-core accretion models of \citet{Mordasini18} are also in good agreement with the overall shape of the semi-major axis distribution, but they underestimate the absolute occurrence rate of giant planets. This could be due to the finite number of cores injected into their simulations.

One common theme among planet formation models of gas giants is that protoplanets tend to migrate inward, all the way to the inner edge of the disk, on timescales much shorter than the gas dissipation timescale. This tends to produce an enhancement of occurrence closer to the star and/or many planets being engulfed by the host star. \citet{Jennings18} attempt to solve this issue by simultaneously modeling the effects of photoevaporation and viscous evolution on the gas disk. They find that, depending on the dominant energy of the evaporating photons, this could clear gaps in the disk that halt Type I migration and creates a pile-up of planets at orbital separations between 0.8--4 au. They showed that this can produce very strong and narrow enhancements near certain orbital separations, but it is conceivable that the shape of the final semi-major axis distribution would actually be driven by the spectral energy distributions of host stars during the early years of their formation.

\citet{Hallatt20} also proposed gap formation in the protoplanetary disk shortly after the formation of gas giant planets as a mechanism to slow or halt migration at preferred orbital separations. Their model requires that the giant planets that form further out in the disk are more massive in order to reproduce the observed enhancements. We expect this to be the case if the dust content of disk envelopes is very low.

The observed enhancement in the occurrence rate of sub-Jovian planets near 1--10 au, seen in Figure \ref{fig:semi_dist_sub}, suggests that the processes that drive the formation and pile-up of planets at those orbital distances also apply to these lower-mass planets. It appears just as likely for a gaseous planet to undergo runaway accretion and grow into a Jovian planet as it is to halt that runaway accretion process early and remain in the sub-Saturn regime.

Unfortunately, it is difficult to extract significant constraints on planet formation models from semi-major axis distributions alone. Future planet catalogs produced by Gaia and The Roman Space Telescope will help to measure the precise shape of the occurrence enhancement around 1 au with planet samples several orders of magnitude larger, but the stellar samples will be different from ours. We plan for future works in this series to analyze the host star metallicity, eccentricity, and multiplicity distributions of our sample, in the hopes of uncovering evidence that discriminates between different planet formation models.

\section{Conclusion}
\label{sec:conclusion}
In this work, we utilize the catalogue of stars, RV-detected planets, and completeness contours from Rosenthal et al. (2021) to measure giant planet occurrence as a function of semi-major axis. We applied a hierarchical Bayesian technique to incorporate measured search completeness and uncertainties in our observations into uncertainties in our occurrence rates. Our results are consistent with previous studies that have found a strong enhancement in the occurrence rates of these planets around 1 au.

We find that the occurrence of planets less massive than Jupiter (30 $\leq$ \msini $\leq$ 300 \mearthe) is enhanced near 1--10 au in concordance with their more massive counterparts. We find that a fall-off in giant planet occurrence at larger orbital distances is favored over models with flat or increasing occurrence, with 2.5 $\sigma$ confidence from our broken power-low model and with 1.5 $\sigma$ confidence from our non-parametric model. Additionally, our occurrence measurements beyond 10 au are consistent with with those derived from direct imaging surveys.

Finally, we lay out the methodology and groundwork for future studies of giant occurrence as a function of planet and host-star properties. With these tools, we plan to study the occurrence rates of giant planets in particular configurations. Paper III in the CLS series will examine the relationship between giant planets and smaller companions, while Paper IV will split our sample into single-giant and multiple-giant systems and investigate the differences and commonalities between these two groups. These undertakings may provide new insight into the formation and evolution of this class of planets that played a crucial role in sculpting the final architecture of our own Solar System.

\facility{Keck:I (HIRES)}

\acknowledgments

We thank Jay Anderson, G\'asp\'ar Bakos, Mike Bottom, John Brewer, Christian Clanton, Jason Curtis, Fei Dai, Steven Giacalone, Sam Grunblatt, Michelle Hill, Lynne Hillenbrand, Rebecca Jensen-Clem, John A.\ Johnson, Chris McCarthy, Sean Mills, Teo Mo\v{c}nik, Ben Montet, Jack Moriarty, Tim Morton, Phil Muirhead, Sebastian Pineda, Nikolai Piskunov, Eugenio Rivera, Julien Spronck, Jonathan Swift, Guillermo Torres, Jeff Valenti, Sharon Wang, Josh Winn, Judah van Zandt, Ming Zhao, and others who contributed to the observations and analysis required for the CLS project.  We acknowledge R.\ P.\ Butler and S.\ S.\ Vogt for many years of contributing to this dataset. This research has made use of the Keck Observatory Archive (KOA), which is operated by the W. M. Keck Observatory and the NASA Exoplanet Science Institute (NExScI), under contract with the National Aeronautics and Space Administration.  We acknowledge RVs stemming from HIRES data in KOA with principal investigators from the LCES collaboration (S.\ S.\ Vogt, R.\ P.\ Butler, and N.\ Haghighipour). 

We gratefully acknowledge the efforts and dedication of the Keck Observatory staff for support of HIRES and remote observing. We are grateful to the time assignment committees of the Caltech, the University of California, the University of Hawaii, NASA, and NOAO for their generous allocations of observing time. Without their long-term commitment to radial velocity monitoring, these planets would likely remain unknown.

We thank Ken and Gloria Levy, who supported the construction of the Levy Spectrometer on the Automated Planet Finder, which was used heavily for this research. We thank the University of California and Google for supporting Lick Observatory, and the UCO staff as well as UCO director Claire Max for their dedicated work scheduling and operating the telescopes of Lick Observatory. G.W.H.\ acknowledges long-term support from NASA, NSF, Tennessee State University, and the State of Tennessee through its Centers of Excellence program. A.W.H.\ acknowledges NSF grant 1753582. H.A.K. acknowledges NSF grant 1555095. P.D.\ gratefully acknowledges support from a National Science Foundation (NSF) Astronomy \& Astrophysics Postdoctoral Fellowship under award AST-1903811. The Center for Exoplanets and Habitable Worlds and the Penn State Extraterrestrial Intelligence Center are supported by the Pennsylvania State University and the Eberly College of Science.

This research has made use of NASA's Astrophysics Data System.
This work has made use of data from the European Space Agency (ESA) mission
{\it Gaia} (\url{cosmos.esa.int/gaia}), processed by the {\it Gaia}
Data Processing and Analysis Consortium (DPAC,
\url{cosmos.esa.int/web/gaia/dpac/consortium}). Funding for the DPAC
has been provided by national institutions, in particular the institutions
participating in the {\it Gaia} Multilateral Agreement.

Finally, the authors wish to recognize and acknowledge the significant cultural role and reverence that the summit of Maunakea has long had within the indigenous Hawaiian community.  We are most fortunate to have the opportunity to conduct observations from this mountain.

\software{All code used in this paper is available at  \url{github.com/California-Planet-Search/rvsearch} and \url{github.com/leerosenthalj/CLSII}. This research makes use of GNU Parallel \citep{Tange11}. We made use of the following publicly available Python modules: \texttt{astropy} \citep{Astropy-Collaboration13}, \texttt{matplotlib} \citep{Hunter07}, \texttt{numpy/scipy} \citep{numpy/scipy}, \texttt{pandas} \citep{pandas}, \texttt{emcee} \citep{DFM13}, and \texttt{RadVel} \citep{Fulton18}.}

\bibliographystyle{aasjournal}
\bibliography{references}

\end{document}